\documentclass[journal]{IEEEtran}
\usepackage[utf8]{inputenc}
\usepackage{enumitem}
\usepackage{amsfonts}
\usepackage{amsmath}
\usepackage{amssymb}
\usepackage{array}
\usepackage{bm}
\usepackage{cases}
\usepackage{color}
\usepackage[dvipsnames]{xcolor} 
 
\usepackage[overload]{empheq}
\usepackage{graphicx}   
\usepackage{makecell}
\usepackage{mathrsfs}
\usepackage{multirow} 
\usepackage{threeparttable} 

\usepackage{physics}    
\usepackage{soul}

\usepackage{subcaption}
\usepackage{xspace}

\usepackage{url}
\usepackage[all]{xy}
\usepackage[font=footnotesize,labelsep=period]{caption}

\newcommand{\PreserveBackslash}[1]{\let\temp=\\#1\let\\=\temp}
\newcolumntype{C}[1]{>{\PreserveBackslash\centering}p{#1}}
\newcolumntype{R}[1]{>{\PreserveBackslash\raggedleft}p{#1}}
\newcolumntype{L}[1]{>{\PreserveBackslash\raggedright}p{#1}}

\usepackage{tikz}   
\usepackage{standalone}
\usepackage{pgfplots}
\usetikzlibrary{shapes, arrows,chains,calc,shapes}
\pgfplotsset{compat = newest}
\pgfdeclarelayer{bg}    
\pgfsetlayers{bg,main}  

\def\insys{\textit{InSys}\xspace}
\def\exsys{\textit{ExSys}\xspace}
\def\ODENet{ODE-Net\xspace}
\def\neuroFA{\textit{Neuro-Reachability}\xspace}
\def\neuroRS{neuro-reachsets\xspace}
\def\neuroRScapital{Neuro-reachsets\xspace}
\def\neuroRSCAPital{Neuro-Reachsets\xspace}

\title{Neuro-Reachability of Networked Microgrids}

\author{Yifan~Zhou,  
        and 
		Peng~Zhang, \IEEEmembership{Senior Member, IEEE}
\thanks{This work has been submitted to IEEE Transactions on Power Systems on September 10, 2020. 
}
\thanks{This work was supported in part by the National Science Foundation under Grant OIA-2040599.
}
\thanks{Y. Zhou and P. Zhang are with the Department of Electrical and Computer Engineering, Stony Brook University, New York 11794-2350, USA (e-mails: yifan.zhou.1@stonybrook.edu, p.zhang@stonybrook.edu).}

}

\begin{document}

\maketitle

\begin{abstract}
A neural ordinary differential equations network (\ODENet)-enabled reachability method (\neuroFA) is devised for the dynamic verification of networked microgrids (NMs) with unidentified subsystems and heterogeneous uncertainties. Three new contributions are presented:
1) An \ODENet-enabled dynamic model discovery approach is devised to construct the data-driven state-space model which preserves the nonlinear and differential structure of the NMs system;
2) A physics-data-integrated (PDI) NMs model is established, which empowers various NM analytics; 
and 3) A conformance-empowered reachability analysis is developed to 
enhance the reliability of the PDI-driven dynamic verification. 
Extensive case studies demonstrate the efficacy of the \ODENet-enabled method in microgrid dynamic model discovery, and the effectiveness of the \neuroFA approach in verifying the NMs dynamics under multiple uncertainties and various operational scenarios. 

\end{abstract}

\section{Introduction}

Networked microgrids (NMs) allow microgrids to support coordinately various smart community functions~\cite{zhang2020networked} and help increase electricity resilience~\cite{feng2019EMPF,wan2020distributed}.
However, two major challenges arise in the dynamic analysis of today's low-inertia NMs~\cite{soni2013improvement}, which prevent NMs from serving as dependable resiliency resources: 
I) Lack of effective analytics to handle the combinatorial explosion in verifying the NMs dynamics under the infinitely many uncertain scenarios \cite{zhou2020reachable}, and
II) Unattainablility of accurate models for each and every microgrid, especially the dynamic models of converters, loads and circuits~\cite{wilding2017turning,xu2012preventive}.

Reachability analysis is a novel method which can provably enclose all dynamic trajectories under uncertain perturbations and large disturbances in NMs~\cite{zhou2020reachable,zhou2019rpf,zhou2020reacheigen}. It prevails over traditional time-domain simulations~\cite{liu2019power} and energy function approaches~\cite{yao2019sufficient,bosetti2017transient} mainly due to the capability of processing infinitely many uncertain scenarios efficiently. Even though reachability analysis is proved to be a promising solution to Challenge I, Challenge II above has been a major obstacle that prevents it from being widely adopted in the planning and operations of NMs.

Learning reliable dynamic models for the unidentified subsystems from measurements, therefore, is of paramount importance for the data-driven NMs dynamic analysis. 
Koopman operator and dynamic mode decomposition are popular approaches to constructing linear approximation of nonlinear systems from data~\cite{korda2018linear,surana2016linear}, whereas they are inefficient to establish nonlinear ODE models for rapidly-fluctuating NMs subject to `random walks' of operating points disturbed by uncertainties~\cite{sinha2019robust}.
Plenty of machine learning approaches have been applied for power system dynamic analysis, either for the time-domain trajectory prediction~\cite{guo2015online} or for the stability classification~\cite{ren2019fully,zheng2017deep}. Nevertheless, discovering dynamic models behind data, which is a long standing open problem, is substantially more important in the sense of providing deep insights of the system dynamics and allowing for formal verifications and control of the system.

The overarching goal of this paper is to establish a data-driven method well suited to discovering the strongly nonlinear NMs dynamics as well as to verifying the  NMs dynamics under uncertainties. To this end,
this paper devises a \neuroFA method. The key innovation is to integrate the neural ordinary differential equations network (ODE-Net) with reachability analysis and conformance theory to allow for a data-driven formal verification of the NMs dynamics under uncertainties. 
The contributions of this work are threefold:
\begin{itemize}
    \item An \ODENet-enabled model discovery method is devised to construct a nonlinear ODE model for the uncertainty-perturbed NMs, which can best preserve the dynamic behaviours of NMs  without assuming a priori any specific dynamic modes. This modeling approach can effectively address the data rich, information poor (DRIP) problem widely existing in today's microgrids.
    \item A physics-data-integrated (PDI) modelling approach is then introduced to combine both physics-based and data-driven ODE models, which enables reachability analysis to verify the PDI-NMs dynamics  incorporating the  hierarchical control of DERs and network transients.
    \item Reachability analysis for NMs is further empowered with a conformance theory as a `feedback' mechanism to further improve the reachset accuracy induced by possible inconformance of the PDI-NMs behaviours compared with the real NMs dynamics.
\end{itemize}
 
The remainder of the paper is organized as follows. 
Section~\ref{sec:ODENet} introduces the \ODENet-enabled dynamic model discovery. 
Section~\ref{sec:model} establishes the NMs model integrated by both physics-based and data-driven subsystems.
Section~\ref{sec:RS} devises the conformance-empowered reachability analysis for the PDI-NMs dynamics. 
Section~\ref{sec:sim} presents case studies on a typical NMs system to validate the \neuroFA method.
Finally, Section~\ref{sec:conclusion} concludes the paper.

\section{\ODENet-Enabled Dynamic Model Discovery for Microgrids} \label{sec:ODENet}

\begin{figure}[!t]
    \centering
    \includegraphics[width=\columnwidth]{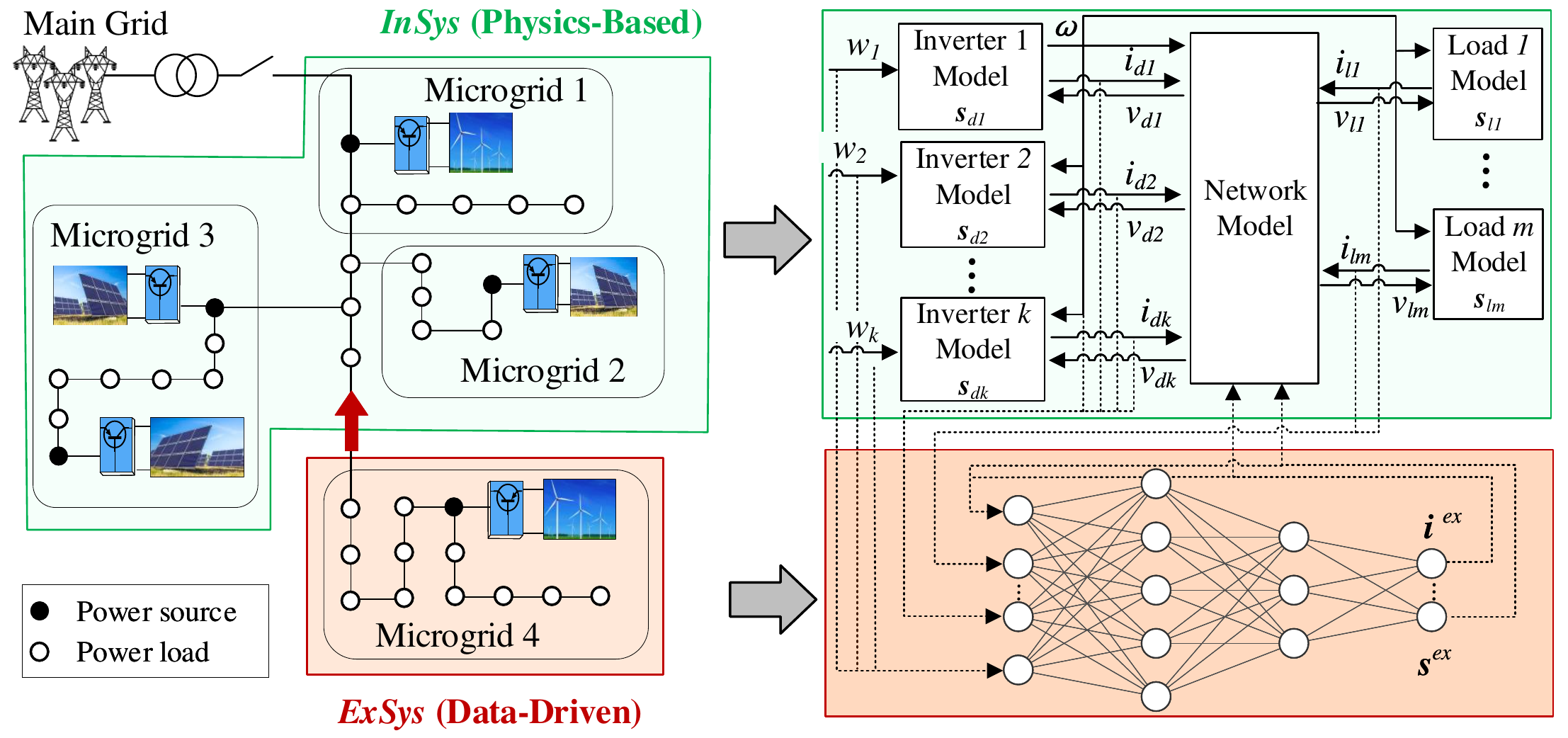}
    \caption{Illustration of NMs modeling with physics-based \insys and data-driven \exsys}
    \label{fig:main:NMs}
    \vspace{-10pt}
\end{figure}

Pursuant to the attainablility of physics models, the overall NMs system can always be partitioned into an internal subsystem (\insys) and an external subsystem (\exsys), as illustrated in Fig.~\ref{fig:main:NMs}. \insys, where the structure and parameters are precisely known, can be readily formulated by assembling the dynamic models of its components. \exsys, in contrast, has to be modeled via a data-driven approach due to the absence of physics models, the unavailability of state measurements, and/or the need to preserve the consumer privacy.

This section devises an \ODENet-enabled method to discover a state-space model of \exsys.

\subsection{\ODENet-Based State-Space Model Formulation} \label{sec:ODENet:formulation}

The functional formulation of \exsys is established as
\begin{equation} \label{equ:ODENet:forward}
    \Dot{\bm{x}} = \mathcal{N}(\bm{x} , \bm{u})
\end{equation}
Here,  function $\mathcal{N}$ represents a state-space form of  \exsys and is to be learned from the measurements; $\bm{x}$ and $\bm{u}$ respectively denote the state variables and input variables of \exsys.
This ODE-governed, learned \exsys model can be integrated with the \insys model for assessing the overall NMs dynamics. Details of $\bm{x}$ and $\bm{u}$ are introduced in Subsection~\ref{sec:model:ExSys}.

\begin{figure}[!t]
    \centering
    \begin{tikzpicture}
            \node[anchor=south west,inner sep=0] at (0,0) {\includegraphics[width=\columnwidth]{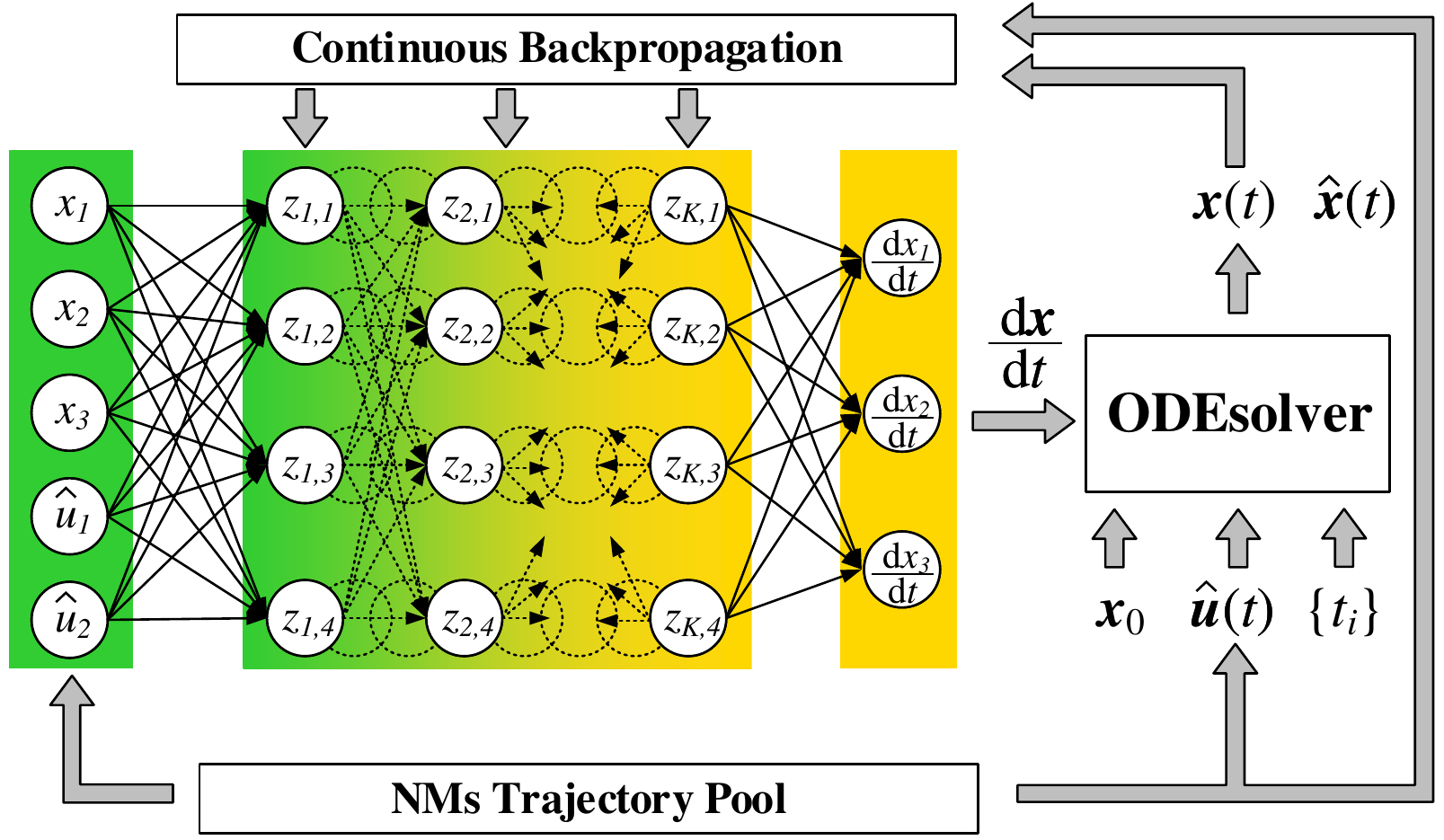}};
            \node at (3.5,0.8) {{$\Dot{\bm{x}} = \mathcal{N}(\bm{x},\bm{u},\bm{\theta}) $}};
        \end{tikzpicture}
    \caption{\ODENet-enabled dynamic model discovery of \exsys}
    \label{fig:main:ODENet}
    \vspace{-10pt}
\end{figure}

Fig.~\ref{fig:main:ODENet} illustrates the dynamic model discovery using \ODENet. 
Taking  $\bm{x}$ and  $\bm{u}$ at time $t$ as the inputs, \ODENet  outputs the time derivative of $\bm{x}$ and therefore explicitly establishes the \exsys dynamic model in \eqref{equ:ODENet:forward} by the forward propagation in the neural network.

Given a time series of NMs trajectories as $\lbrace (t_1, t_2, \dots ,t_n), (\hat{\bm{x}}_1, \hat{\bm{x}}_2, \dots ,\hat{\bm{x}}_n), (\hat{\bm{u}}_1, \hat{\bm{u}}_2, \dots ,\hat{\bm{u}}_n) \rbrace$, \ODENet best matches the \exsys dynamics by minimizing the error between the state measurements $\hat{\bm{x}}$ and the numerical solution of \eqref{equ:ODENet:forward}:
\begin{equation} \label{equ:ODENet:loss min}
\begin{aligned}
   & \min_{\bm{\theta}} \sum_{i=1}^n  L(\bm{x}_i) = \sum_{i=1}^n \frac{1}{2} \eta_i  \| \bm{x}_i -  \hat{\bm{x}}_i \|_2 \\
   & s.t.~~ \bm{x}_i =   \hat{\bm{x}}_1 + \int_{t_1}^{t_i} \mathcal{N}(\bm{x},\bm{u},\bm{\theta})\dd t 
\end{aligned}
\end{equation}
where $\bm{\theta}$ denotes the \ODENet parameters; $\eta_i$ denotes the weighting factor at time point $i$.

\subsection{Continuous Backpropagation Technique} \label{sec:ODENet:loss}
The main difficulty in the optimization of \eqref{equ:ODENet:loss min} lies in the ODE integration operation in the constraints.
In this subsection, the continuous propagation technique~\cite{chen2018neural} is applied to handle the ODE integration in the \ODENet training.

Lagrange multiplier $\bm{\lambda}$ is first introduced to \eqref{equ:ODENet:loss min} to  remove the ODE constraints and build the following loss function: 
\begin{equation}\label{equ:ODENet:loss}
    \mathcal{L} =  \sum_{i=1}^n L(\bm{x}_i) - \int_{t_0}^{t_n} \bm{\lambda}^T \left(\Dot{\bm{x}} - \mathcal{N}(\bm{x},\bm{u}, \bm{\theta}) \right) \dd t
\end{equation}

Backpropagation computes the gradient of the loss function  with respect to the \ODENet parameters to minimize the loss function ~\cite{goodfellow2016deep}. With the loss function \eqref{equ:ODENet:loss} involving the integration operator, the partial derivative of $\mathcal{L}$ with respect to  $\bm{\theta}$ is calculated as:

\begin{small}
\begin{equation}
\begin{aligned}
    \pdv{\mathcal{L}}{\bm{\theta}} = 
   & \sum_{i=1}^n \left( \pdv{ L }{ \bm{x}_i } \pdv{ \bm{x}_i }{ \bm{\theta} }  - \int_{t_{i-1}}^{t_i} \bm{\lambda}^T \left(\pdv{\Dot{\bm{x}}}{\bm{\theta}} - \pdv{ \mathcal{N} }{ \bm{x} } \pdv{ \bm{x} }{ \bm{\theta} } - \pdv{ \mathcal{N} }{ \bm{\theta} } \right) \dd t \right)\\
   = & \sum_{i=1}^n \pdv{ L }{ \bm{x}_i } \pdv{ \bm{x}_i }{ \bm{\theta} }  
    + \sum_{i=1}^n  \left( \bm{\lambda}^T(t_i^-) \pdv{ \bm{x}_i }{ \bm{\theta} }  - \bm{\lambda}^T(t_{i-1}^+) \pdv{ \bm{x}_{i-1} }{ \bm{\theta} }  \right) \\
   & + \sum_{i=1}^n \int_{t_{i-1}}^{t_i} \left(  
        \dv{\bm{\lambda}^T}{t}   \pdv{\bm{x}}{\bm{\theta}} 
        + \bm{\lambda}^T \pdv{ \mathcal{N} }{ \bm{x} } \pdv{ \bm{x} }{ \bm{\theta} } 
        +  \bm{\lambda}^T \pdv{ \mathcal{N} }{ \bm{\theta} } \right) \dd t \nonumber
\end{aligned}
\end{equation}
\end{small}
\begin{equation}\label{equ:ODENet:pdfL}
    ~~
\end{equation}

The Lagrange multiplier variables are given by~\cite{sun2019neupde}:
\begin{equation}\label{equ:ODENet:lambda}
    \dv{\bm{\lambda}^T}{t} = - \bm{\lambda}^T \pdv{ \mathcal{N} }{ \bm{x} }
\end{equation}
where the boundary conditions are set as:
\begin{equation}\label{equ:ODENet:lambda boundry}
    \bm{\lambda}^T(t_n^+) = 0 ~~,~~ \bm{\lambda}^T(t_i^+) = \bm{\lambda}^T(t_i^-) + \pdv*{ L }{ \bm{x}_i }
\end{equation}

Then, \eqref{equ:ODENet:pdfL} can be derived into:
\begin{equation}\label{equ:ODENet:pdfL2}
\pdv{\mathcal{L}}{\bm{\theta}}
   = \int_{t_{0}}^{t_n}  \bm{\lambda}^T \pdv{ \mathcal{N} }{ \bm{\theta} }  \dd t 
\end{equation}

Collecting \eqref{equ:ODENet:lambda} and \eqref{equ:ODENet:pdfL2} leads to an ODE integration problem:
\begin{equation}\label{equ:ODENet:Ltheta}
    \dv{}{t} 
    \begin{bmatrix} \bm{\lambda}^T  \\ \pdv*{\mathcal{L}}{\bm{\theta}} \end{bmatrix} 
    = \begin{bmatrix}  
    - \bm{\lambda}^T \pdv*{ \mathcal{N} }{ \bm{x} } \\
    \bm{\lambda}^T  \pdv*{ \mathcal{N} }{ \bm{\theta} }
    \end{bmatrix} 
\end{equation}

Subsequently, $\pdv*{\mathcal{L}}{\bm{\theta}}$ can be obtained from \eqref{equ:ODENet:Ltheta} by any ODE solver, e.g., Trapezoidal integration. Given the final value of $\lambda(t)$, i.e., $\bm{\lambda}^T(t_n)$ in \eqref{equ:ODENet:lambda boundry}, rather than the initial value, \eqref{equ:ODENet:Ltheta} requires solving the ODEs backwards in time, which leads to a reverse-mode integration~\cite{chen2018neural}:

\begin{small}
\begin{equation}\label{equ:ODENet:L}
    \eval{\pdv{\mathcal{L}}{\bm{\theta}}}_{t_1} = \eval{\pdv{\mathcal{L}}{\bm{\theta}}}_{t_n} + \int_{t_{n}}^{t_{1}} \bm{\lambda}^T  \pdv{ \mathcal{N} }{ \bm{\theta} } \dd t 
    = \sum_{i=2}^n \int_{t_{i}}^{t_{i-1}} \bm{\lambda}^T  \pdv{ \mathcal{N} }{ \bm{\theta} } \dd t
\end{equation}
\end{small}
with $\bm{\lambda}(t)$ also solved by the reverse-mode integration:
\begin{equation}
    \bm{\lambda}^T (t_{i-1}^+) = \bm{\lambda}^T (t_{i}^-) + \int_{t_{i}}^{t_{i-1}}  \bm{\lambda}^T \pdv{ \mathcal{N} }{ \bm{x} }\dd t
\end{equation}

Further, consider \textit{a set of time series} of NMs trajectories as $\lbrace (\bm{t}^{(1)}, \hat{\bm{x}}^{(1)}, \hat{\bm{u}}^{(1)}),\dots, (\bm{t}^{(m)}, \hat{\bm{x}}^{(m)}, \hat{\bm{u}}^{(m)}) \rbrace$.  For the $j^{th}$ measurement, let $\pdv*{\mathcal{L}^{(j)}}{\bm{\theta}} $ be the gradient of the loss function computed by \eqref{equ:ODENet:L}.
The overall gradient is obtained:
\begin{equation}\label{equ:ODENet:reverse 1}
    \pdv{\mathcal{L}}{\bm{\theta}} = \sum\nolimits_{j = 1}^m \pdv{\mathcal{L}^{(j)}}{\bm{\theta}} 
\end{equation}

Consequently, the \ODENet parameters are updated using gradient descent so that $\mathcal{L}$ can be decreased during training:
\begin{equation}\label{equ:ODENet:reverse 2}
     \bm{\theta} \longleftarrow \bm{\theta} - r \pdv{\mathcal{L}}{\bm{\theta}}
\end{equation}
where $r$ denotes the learning rate.

The continuous backpropagation incorporates the ``ODE solver'' in the gradient descent for the \ODENet parameter optimization, and hence effectively retains the intrinsic continuous differential structure of the dynamical NMs.

\subsection{Continuous Network Structure} \label{sec:ODENet:structure}
Despite the capability of modeling complicated nonlinear dynamics, a deep \ODENet involves a large number of parameters and prohibitively high cost for computing $\pdv*{\mathcal{N}}{ \bm{x} }$ and  $\pdv*{ \mathcal{N} }{ \bm{\theta} }$ in the continuous backpropagation. 
 To resolve this problem, this paper uses a \emph{continuous-depth network}, which enables a linear memory cost with network depth and controllable numerical error~\cite{chen2018neural}.

In a classical neural network, the hidden layers follow a  discrete structure (i.e., layers 1, 2, $\cdots$, $K$ as illustrated by the solid-line neurons in Fig.~\ref{fig:main:ODENet}), and therefore formulates a set of difference equations for forward propagation:
\begin{equation}\label{equ:ODENet:discrete}
    \bm{z}_{k+1} = \bm{z}_{k} + \bm{h}( \bm{z}_{k} ,  \bm{\theta}_{k})
\end{equation}
Here, $k \in \mathbb{N}^+$ denotes the discrete layer; $\bm{z}_{k}$ and $\bm{\theta}_{k}$ respectively denote the output states and parameters of the $k^{th}$  hidden layer.

The continuous-depth network regards the forward propagation of the discrete layers in \eqref{equ:ODENet:discrete} as an Euler discretization  of a set of continuous differential equations, i.e., it continuously propagates the states from the input layer to the output layer (see the dotted-line neurons in Fig.~\ref{fig:main:ODENet} ). This idea leads to  a  continuous ``layer dynamics'': 
\begin{equation}\label{equ:ODENet:continuous}
    \dv{\bm{z}(k)}{k} = \bm{h}( \bm{z}(k) ,  \bm{\theta}_h)
    \Longrightarrow
    \bm{z}_K = \bm{z}_1 {+} \int_{1}^{K}\bm{h}( \bm{z},  \bm{\theta}_h)  \dd k
\end{equation}
Here, $k \in \mathbb{R}^+$ denotes the continuous layer; $\bm{z}(k)$ and $\bm{\theta}_h$ respectively denote the output states and parameters of the continuous hidden layer.
As can be seen from \eqref{equ:ODENet:continuous}, $\bm{z}_K$, i.e., the final output of the hidden layer, can be directly obtained by integrating over the continuous layers.

Based on the continuous-depth network, the chain rule is applied to compute $\pdv*{\mathcal{N}}{ \bm{x} }$ and $\pdv*{\mathcal{N}}{ \bm{x} } $ for \eqref{equ:ODENet:Ltheta}:
\begin{equation}
    \pdv{\mathcal{N}}{ \bm{x} }  = \bm{\theta}_K^T \pdv{\bm{z}_K}{\bm{z}_1}\pdv{\bm{z}_1}{\bm{x}}
    ~~,~~
    \pdv{\mathcal{N}}{ \bm{\theta}_h } = \bm{\theta}_K^T \pdv{\bm{z}_K}{ \bm{\theta}_h }
\end{equation}
To obtain $\pdv*{\bm{z}_K}{\bm{z}_1}$ and $\pdv*{\bm{z}_K}{ \bm{\theta}_h }$ in the equation above, the reverse-mode ODE integration discussed in Subsection~\ref{sec:ODENet:loss} is again applied. Specifically, in analogy to \eqref{equ:ODENet:loss}, the following ODEs are formulated~\cite{chen2018neural} for the gradient of $\bm{z}_K$ with respect to $\bm{z}(k)$:
\begin{equation} \label{equ:ODENet:reverse 3}
    \dv{}{k} 
    \begin{bmatrix} \pdv*{\bm{z}_K}{\bm{z}}  \\ \pdv*{\bm{z}_K}{\bm{\theta}_h} \end{bmatrix} 
    = \begin{bmatrix}  
    - (\pdv*{\bm{z}_K}{\bm{z}})^T \pdv*{\bm{h}}{\bm{z}} \\
    (\pdv*{\bm{z}_K}{\bm{z}})^T \pdv*{\bm{h}}{\bm{\theta}_h }
    \end{bmatrix} 
\end{equation}
Then, in analogy to \eqref{equ:ODENet:reverse 1} and \eqref{equ:ODENet:reverse 2}, the numerical solution of \eqref{equ:ODENet:reverse 3}, i.e., $\pdv*{\bm{z}_K}{\bm{z}_1}$ and $\pdv*{\bm{z}_K}{ \bm{\theta}_h }$, is obtained by the ODE integration.

\section{Physics-Data-Integrated ODE Modeling for NMs Dynamics} \label{sec:model}

Based on the \ODENet-enabled \exsys formulation, this section establishes the overall dynamic model of NMs by combining the physics-based \insys and data-driven \exsys, as illustrated in Fig.~\ref{fig:main:NMs}.

\subsection{Physics-Based Formulation of \insys} \label{sec:model:InSys}

Given the model of each component (e.g., a DER, a power load, a branch), \insys can be explicitly formulated as a system of differential algebraic equations (DAEs):
\begin{subequations}  \label{equ:InSys:DAE}
    \begin{align}
        &  \Dot{\bm{s}}^{in} = \bm{g}^{in}(\bm{s}^{in},\bm{i}^{in},\bm{w}^{in}, \bm{s}^{ex}) \label{equ:InSys:DAE1}\\
        &  \Dot{\bm{i}}^{in} =  \bm{f}^{in}(\bm{s}^{in},\bm{i}^{in}) + \bm{m}^{in}\bm{v} \label{equ:InSys:DAE2}\\
        & \bm{n}^{in} \bm{i}^{in}+\bm{n}^{ex} \bm{i}^{ex} = \bm{0}  \label{equ:InSys:DAE3}
    \end{align}
\end{subequations}
Here, \eqref{equ:InSys:DAE1} formulates the dynamics of each component in \insys, where $\bm{s}^{in}$ denotes the state variables of each component; 
$\bm{i}^{in}$ denotes the current injections from each component and is normally formulated under the DQ coordinates for the inverter-dominated NMs; 
$\bm{w}^{in}$ denotes the uncertain inputs caused by DERs in \insys;   
$\bm{s}^{ex}$ includes (but not limited to) the control variables sent from \exsys, which is particularly used to account for the global control effects;
function $\bm{g}^{in}$ is formulated according to the dynamics of each component.
Equation \eqref{equ:InSys:DAE2} formulates the current injections from each component, where $\bm{v}$ denotes the DQ-axis voltages at each bus of \insys and boundary buses of \exsys; $\bm{m}^{in}$ denotes the DQ-axis incidence matrix between components and buses.
Equation \eqref{equ:InSys:DAE3} formulates the Kirchhoff's Current Law at the buses of \insys and boundary buses of \exsys, where $\bm{i}^{ex}$ denotes the current injections from \exsys; matrices $\bm{n}^{in}$ and $\bm{n}^{ex}$  denotes the directed component-bus incidence matrix.

\subsection{Data-Driven Formulation of \exsys}\label{sec:model:ExSys}

Following \eqref{equ:InSys:DAE}, \exsys interacts with \insys through its current injections $\bm{i}_{ex}$ and control signals $\bm{s}_{ex}$. Therefore, $\bm{i}_{ex}$ and $\bm{s}_{ex}$ are retained for dynamic simulations of \insys. 
Accordingly, \exsys is formulated as:
\begin{subequations} \label{equ:ExSys}
\begin{align}
    & \Dot{\bm{s}}^{ex} = \bm{g}^{ex}( \bm{i}^{in}, \bm{s}^{in}, \bm{i}^{ex}, \bm{s}^{ex}, \bm{w}^{in}, \bm{w}^{ex}) \label{equ:ExSys1}\\
    & \Dot{\bm{i}}^{ex} = \bm{f}^{ex}( \bm{i}^{in}, \bm{s}^{in}, \bm{i}^{ex}, \bm{s}^{ex}, \bm{w}^{in}, \bm{w}^{ex}) \label{equ:ExSys2}
\end{align}
\end{subequations}
Here, \eqref{equ:ExSys1} formulates the dynamics of control variables sending from \exsys to \insys; \eqref{equ:ExSys2} formulates the dynamics of the current injections from \exsys; $\bm{w}^{ex}$ denotes the uncertain factors in \exsys.

\ODENet is used to establish the ODE model in \eqref{equ:ExSys}, following the procedures in Section~\ref{sec:ODENet}. 
The \exsys formulation in \eqref{equ:ODENet:forward} can now be expanded by: $\mathcal{N} = [\bm{g}^{ex};\bm{f}^{ex}]$ denoting the dynamics of \exsys;
$\bm{x} = [\bm{s}^{ex} ; \bm{i}^{ex}]$ denoting the state variables of \exsys; and $\bm{u} = [\bm{i}^{in}; \bm{s}^{in}; \bm{w}^{in}; \bm{w}^{ex}]$ assembling the input variables in \eqref{equ:ExSys}.

\subsection{Physics-Data-Integrated (PDI)  NMs Model}\label{sec:model:integration}
The entire NMs model is established by combining the physics-based formulation of \insys and data-driven formulation of \exsys:
\begin{subequations} \label{equ:NMs:DAE}
    \begin{align}[left = \empheqlbrace\,]
        &  \Dot{\bm{s}} = \bm{g}(\bm{i},\bm{s},\bm{w})  \label{equ:NMs:DAE 1}\\
        &  \Dot{\bm{i}} = \bm{f}(\bm{i},\bm{s},\bm{w})  + \bm{mv} \label{equ:NMs:DAE 2}\\
        &   \bm{0} = \bm{ni}  \label{equ:NMs:DAE 3}
    \end{align}
\end{subequations}
where $\bm{s}$, $\bm{i}$ and $\bm{w}$ respectively assemble the state variables, current injections and uncertainties of \insys and \exsys.

The  DAE model in \eqref{equ:NMs:DAE} can be rigorously converted to a system of nonlinear ODEs~\cite{zhou2020ode}, as briefly introduced below.
Denote the matrix constructed by the maximal linearly independent columns of $\bm{n}$ as $\bm{n}_1$, and the matrix constructed by the other columns as $\bm{n}_0$. Since $\bm{n}_1$ is non-singular,  \eqref{equ:NMs:DAE 3} leads to the following:
\begin{equation} \label{equ:NMs:x1}
    \bm{n}_0 \bm{i}_0 + \bm{n}_1 \bm{i}_1 = \bm{0} 
    \Longrightarrow
    \bm{i}_1 =  - \bm{n}_1^{-1}\bm{n}_0 \bm{i}_0
\end{equation}
where $\bm{i}_0$ and $\bm{i}_1$ respectively denote the sub-vectors of $\bm{i}$ corresponding to $\bm{n}_0$ and $\bm{n}_1$.

Taking derivative of \eqref{equ:NMs:DAE 3} yields the following: 
\begin{equation}
  \bm{0} =  \bm{n}  \Dot{\bm{i}}  =  \bm{n} \hat{\bm{f}}(\bm{i}_0,\bm{s}) + \bm{nm} \bm{v} 
  \Longrightarrow
  \bm{v}  = - (\bm{nm} )^{-1}\bm{n} \hat{\bm{f}} 
\end{equation}
where $\hat{\bm{f}}(\bm{i}_0,\bm{s},\bm{w}) = \bm{f}(\bm{i},\bm{s},\bm{w})$ by substituting \eqref{equ:NMs:x1} to $\bm{f}$.

Therefore, \eqref{equ:NMs:DAE} is converted to an ODE model:
\begin{subequations} \label{equ:NMs:ODE}
    \begin{align}[left = \empheqlbrace\,]
        &  \Dot{\bm{s}} = \hat{\bm{g}}(\bm{i}_0,\bm{s},\bm{w}) \\
        &  \Dot{\bm{i}}_0 = -\bm{m}_0 (\bm{nm})^{-1} \bm{n} \hat{\bm{f}}(\bm{i}_0,\bm{s}) + \hat{\bm{f}}_0(\bm{i}_0,\bm{s})
    \end{align}
\end{subequations}
where $\bm{m}_0$ and $\hat{\bm{f}}_0$ respectively extract the components of $\bm{m}$ and $\hat{\bm{f}}$ corresponding to $\bm{i}_0$.

The obtained ODE model in \eqref{equ:NMs:ODE} is rigorously equivalent to the original DAE model \eqref{equ:NMs:DAE 1} without adopting any linearization. Hence, it can be used for the transient analysis under small or large disturbances.
The PDI-NMs model in \eqref{equ:NMs:ODE} can then be abstracted as:
\begin{equation} \label{equ:NMs}
    \Dot{\bm{X}} = \bm{F}(\bm{X} , \bm{W})
\end{equation}
where $\bm{X}$ denotes the state variables of NMs integrating states of both \insys and \exsys; $\bm{W}$ denotes to the uncertainty inputs.

\section{Neuro-Reachability Analysis of NMs Dynamics } \label{sec:RS}

Based on the PDI-NMs model, this section devises a \neuroFA method for dynamic verification of NMs.

\subsection{Reachset For NMs} \label{sec:RS:reachset}
Reachability analysis verifies the NMs dynamics by calculating reachsets, i.e., a provable enclosure of all possible dynamic trajectories under infinitely many uncertain scenarios. 
Given the set of initial NMs states $\mathcal{X}^0$ and the set of DER uncertainties $\mathcal{W}$, the time-point reachset is defined as the set of all the possible NMs states at time $t$: 
\begin{equation}\label{equ:rs:tp def}
\begin{aligned}
    \mathcal{R} ( t ) =&  \Bigl\lbrace  \bm{X}( t ) =  \int_0^t \bm{F}(\bm{X}( \tau ),\bm{W}( \tau )) \dd \tau   \Bigm|\\
    & ~~  \bm{X}(0)\in \mathcal{X}^0, \bm{W} \in \mathcal{W} \Bigr\rbrace
\end{aligned}
\end{equation}
In this research, the uncertainty set $\mathcal{W}$ is formulated by an unknown-but-bounded set; and zonotope bundle is utilized for set formulation~\cite{althoff2011zonotope}, i.e., an intersection of finite zonotopes which is able to represent arbitrary polytopes with a satisfactory computational efficiency.

Applying Taylor expansion to \eqref{equ:NMs} at $(\bm{X}^*, \bm{W}^*)$ gives:
\begin{equation}\label{equ:rs:lin}
  \begin{aligned}
        \Dot{\bm{X}}  \in \bm{F}^* + \bm{J}(\bm{X}-\bm{X}^*) + \bm{J}_w (\bm{W}-\bm{W}^*) + \bm{e}_l
  \end{aligned}
\end{equation}
where   $\bm{F}^* = \bm{F}(\bm{X}^*,\bm{W}^*)$; $\bm{J}= \pdv*{\bm{F}^*}{\bm{X}} $ and $\bm{J}_w = \pdv*{\bm{F}^*}{\bm{W}} $ respectively denote the Jacobian matrix referring to the first-order Taylor series; $\bm{e}_l$ is the Lagrange remainder~\cite{althoff2008nonlinear}.
For computational convenience,  $\mathcal{R}(t)$ is shifted to  $\mathcal{R}(t)-\bm{X}^*$ to perform the set computation.
Therefore, the time-point reachset of the NMs dynamics can be computed by the evolution of the previous time-point reachset:
\begin{equation} \label{equ:rs:R point}
    \mathcal{R}(t)  = \left( e^{\bm{J}\Delta} \mathcal{R}(t-\Delta) \right) \oplus \int_0^{\Delta} e^{\bm{J}(\Delta-\tau)} \mathcal{W}_F\dd \tau \oplus \mathcal{R}_{el}
\end{equation}
where $\Delta$ denotes the time step; $\oplus$ denotes the Minkowski addition.
In \eqref{equ:rs:R point}, the first term computes the reachset propagation resulting from the NMs states, i.e., $\bm{J}(\bm{X}-\bm{X}^*)$.
The second term computes the reachset propagation resulting from the inputs $\mathcal{W}_F = \bm{J}_w(\mathcal{W}-\bm{W}^*) +\bm{F}^*$, which composes both the uncertainty impact $\bm{J}_w(\bm{W}-\bm{W}^*)$ and the linearization point impact $\bm{F}^*$.
And the third term in \eqref{equ:rs:R point} computes the set of linearization error~\cite{althoff2008nonlinear} to ensure an provable over-approximation of the nonlinear NMs dynamics. 

Further, the reachset during time interval $[t-\Delta,t]$ can be calculated as the union of the time-point reachable sets during the interval as $\mathcal{R}([t-\Delta,t]) = \cup_{\tau \in [t-\Delta,t]}  \mathcal{R} ( \tau )$.

\subsection{Conformance Reachset For PDI-NMs}\label{sec:RS:conformance}

The ODE-Net-enabled model learned from a finite set of training samples, although sufficiently precise for the training set, would not perfectly replicate the real dynamics of the system under any circumstances.
To address the possible inaccuracy of the PDI-NMs model, this subsection empowers the reachability analysis done in Subsection~\ref{sec:RS:reachset} with the conformance theory as a `feedback' mechanism to further improve the \neuroFA reliability.

\subsubsection{Conformance-Empowered Reachset Formulation}

Incorporating the model inaccuracy into the Taylor expansion of NMs dynamics leads to the following:
\begin{equation}\label{equ:rs conf:lin}
   \bm{\Dot{X}} \in \bm{F}^* + \bm{J}  (\bm{X}-\bm{X}^*)+\bm{J}_w (\bm{W}-\bm{W}^*)  + \bm{e}_l + \bm{e}_m
\end{equation} 
Compared with \eqref{equ:rs:lin}, an additional term $\bm{e}_m$ is introduced to address the impact of the discrepancy between the PDI-NMs model and the real NMs trajectories.
In this research, the PDI-NMs model inaccuracy is formulated as multidimensional intervals $\mathcal{E}_m = [\underline{\bm{e}}_m , \overline{\bm{e}}_m]$, where $\underline{\bm{e}}_m $ and $\overline{\bm{e}}_m$ respectively denotes the infimum and supremum of the PDI-NMs model inaccuracy.
Correspondingly, the reachset computation can be modified as:
\begin{equation}\label{equ:rsc:Rc}
\begin{aligned}
    \mathcal{R}_{c}(t)  = & \left( e^{\bm{J}\Delta} \mathcal{R}_c(t-\Delta)  \right) \oplus \mathcal{R}_{el} \oplus \\
     &  \int_0^{\Delta} e^{\bm{J}(\Delta-\tau)}(\mathcal{W}_F+\mathcal{E}_m) \dd \tau
\end{aligned}
\end{equation}
Here, $\mathcal{R}_{c}$ denotes a conformance reachset incorporating the PDI-NMs model error $\mathcal{E}_m$.

Given a specific $\mathcal{E}_m$, $\mathcal{R}_{c}$ can be readily computed following Subsection~\ref{sec:RS:reachset}.
However, theoretically, $\mathcal{E}_m$ is a posteriori error depending on specific NMs states and DER uncertainties, and is supposed to be computed by comparing the PDI-NMs dynamics with the real NMs trajectories.  
To tackle this difficulty, an optimization approach to estimating $\mathcal{E}_m$ is introduced by constructing a minimal-volume conformance reachset while ensuring a provable enclosure of the time-series measurements of the real NMs dynamics.

\subsubsection{$\mathcal{E}_m$-Optimization Model}

Given a time series of NMs trajectories 
$\lbrace \bm{t}, \hat{\bm{X}}, \hat{\bm{W}} \rbrace = 
\lbrace (t_1, t_2, \dots ,t_n), (\hat{\bm{X}}_1, \hat{\bm{X}}_2, \dots ,\hat{\bm{X}}_n), (\hat{\bm{W}}_1, \hat{\bm{W}}_2, \dots ,\hat{\bm{W}}_n) \rbrace$, $\mathcal{R}(t)$ can be computed following Subsection~\ref{sec:RS:reachset} with the NMs initial state set as $\mathcal{X}^0 = \{ \hat{\bm{X}}_1 \}$ and the DER uncertainty set as $\mathcal{W} = \{ \hat{\bm{W}} \}$. 
As discussed above, since the PDI-NMs model may not perfectly conform with the real NMs dynamics, the reachset $\mathcal{R}(t)$ from \eqref{equ:rs:R point}, computed with $\mathcal{W}$ and $\mathcal{X}^0$,  possibly does not enclose $\hat{\bm{X}}$.
Inspired by the conformance theory in~\cite{liu2018reachset,kochdumper2020establishing}, $\mathcal{E}_m$ is optimized to ensure that the conformance reachset  $\mathcal{R}_c(t)$ encloses the time-series NMs states $\hat{\bm{X}}$:
\begin{equation}\label{equ:rsc:opt}
\begin{aligned}
   \mathcal{E}_m \text{-\textbf{opt model}: } &\min_{\mathcal{E}_m} \sum\nolimits_{t \in \bm{t}} \text{Vol}(\mathcal{R}_{c}(t)) \\
    & s.t. ~~ \hat{\bm{X}}(t) \in \mathcal{R}_{c}(t)~~,~~\forall t \in \bm{t}
\end{aligned}
\end{equation}
Here, $\mathcal{R}_{c}$ is computed as \eqref{equ:rsc:Rc} with aforementioned $\mathcal{X}^0$ and $\mathcal{W}$; $\text{Vol}(\cdot)$ computes the volume of the reachset.
Optimization in \eqref{equ:rsc:opt} solves an $\mathcal{E}_m$ such that  $\mathcal{R}_{c}$  encloses $\hat{\bm{X}}$ with a minimal modification on the reachset. As a special case, if the PDI-NMs model perfectly replicates the NMs trajectories $\hat{\bm{X}}$, \eqref{equ:rsc:opt} gives $\mathcal{E}_m = \emptyset$.

\begin{figure}[!t]
    \centering
    \begin{tikzpicture}
    \node[anchor=south west,inner sep=0] at (0,-0.4)
    {\includegraphics[width=252pt]{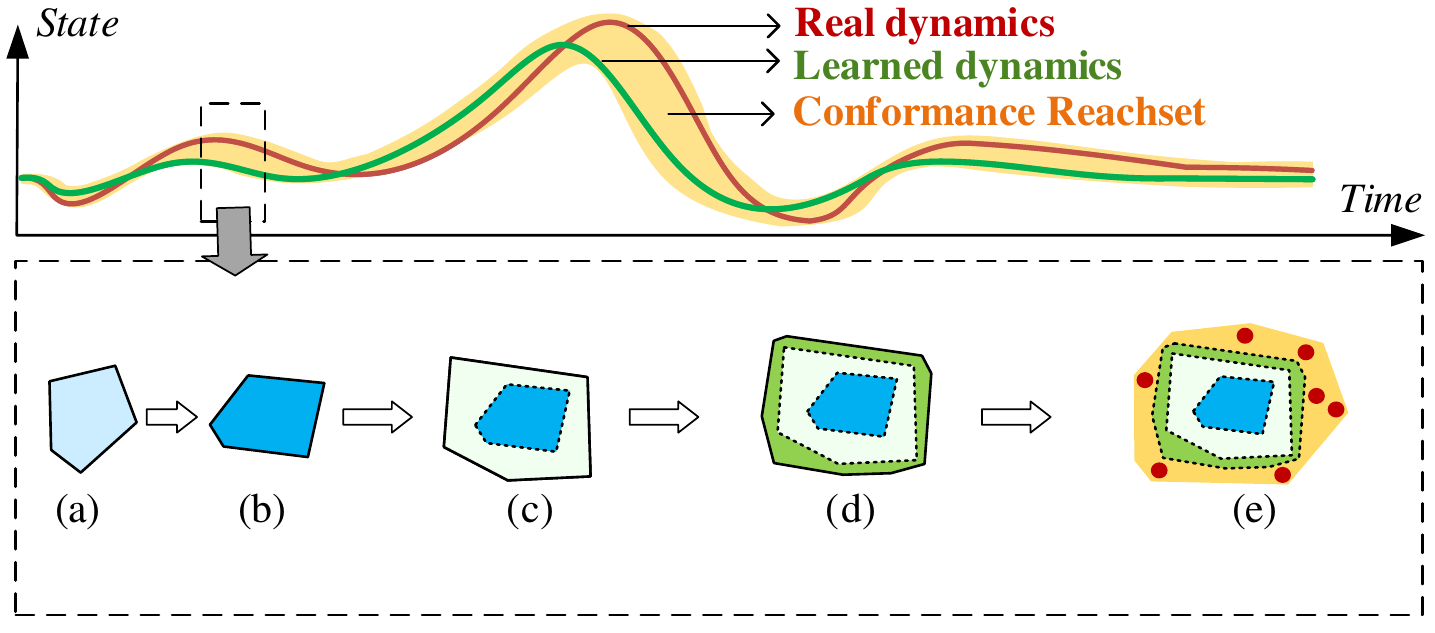}};
    \node at (1.5,3.1) {\scriptsize \textcolor{black}{$[t{-}\Delta,t]$}};
    \node at (0.75,1.55) {\scriptsize \textcolor{black}{$\mathcal{R}(t{-}\Delta)$}};
    \draw[->, line width=0.5](0.6,1.0) -- (0.6,1.4);
    \node at (2.2,1.55) {\scriptsize \textcolor{black}{$e^{\bm{J} \Delta} \mathcal{R}(t{-}\Delta)$}};
    \draw[->, line width=0.5](1.7,1.0) -- (2,1.4);
    \node at (2.3,0) {\scriptsize \textcolor{LimeGreen}{Input $\int_0^{\Delta} e^{\bm{J}(\Delta-\tau)} \mathcal{W}_F\dd \tau$}};
    \draw[->, line width=0.5](3.2,0.65) -- (2.8,0.3);
    \node at (4.8,0.1) {\scriptsize \textcolor{OliveGreen}{Linearization}};
    \node at (4.65,-0.12) {\scriptsize \textcolor{OliveGreen}{error $\mathcal{R}_{el}$}};
    \draw[->, line width=0.5](4.9,0.75) -- (4.8,0.2);
    \node at (5.3,1.7) {\small \textcolor{OliveGreen}{$\bm{\mathcal{R}(t)}$}};
    \node at (6.5,0.1) {\scriptsize \textcolor{purple}{Real trajectory}};
    \node at (6.5,-0.12) {\scriptsize \textcolor{purple}{measurements}};
    \draw[->, line width=0.5](7.15,0.6) -- (6.8,0.25);
    \node at (8,0.1) {\scriptsize \textcolor{BurntOrange}{ODE-Net}};
    \node at (8.0,-0.12) {\scriptsize \textcolor{BurntOrange}{error}};
    \draw[->, line width=0.5](8.15,0.85) -- (8.2,0.3);
    \node at (7.7,1.7) {\small \textcolor{OrangeRed}{$\bm{\mathcal{R}_c(t)}$}};
    \end{tikzpicture}
    \caption{Illustration of conformance reachset}
    \label{fig:main:conformance}
\end{figure}
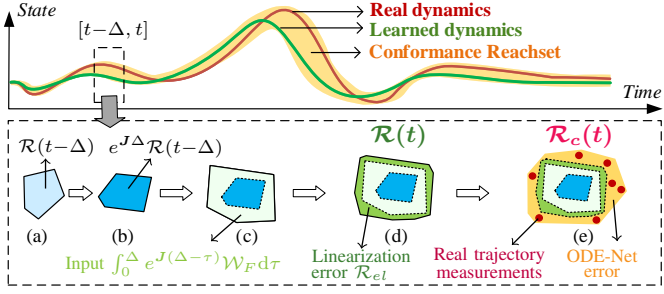

Fig.~\ref{fig:main:conformance} illustrates the basic idea of this optimization-based conformance reachset method. For better visualization, the trajectories generated by the PDI-NMs model (denoted by the green line) differs largely from the real NMs trajectories (denoted by the red line), which further leads to a quite loose conformance reachset (denoted by the yellow area). Fortunately, case studies in Section~\ref{sec:sim} will show that the \ODENet-based dynamic model is quite accurate and the obtained \neuroRS is rather tight.
Taking time period $[ t-\Delta , t ]$  as an example, following \eqref{equ:rs:R point}, the conventional reachset $\mathcal{R}(t)$ (Fig.~\ref{fig:main:conformance}(d)) is computed incorporating the evolution of NMs states (Fig.~\ref{fig:main:conformance}(b)), DER uncertainty inputs (Fig.~\ref{fig:main:conformance}(c)) and linearization error (Fig.~\ref{fig:main:conformance}(d)). Then, as illustrated in Fig.~\ref{fig:main:conformance}(e),  the conformance reachset $\mathcal{R}_c(t)$, which amends $\mathcal{R}(t)$ with the impact of the model inaccuracy set $\mathcal{E}_m$, will enclose the real NMs states at time $t$  with a minimized set volume.
The above process is successively computed over the time horizon to optimize a $\mathcal{E}_m$ such that $\mathcal{R}_c$ encloses the real NMs trajectories.

The $\mathcal{E}_m$-opt model  is intractable to optimize due to the complicated set calculations in the objective and constraints. 
For efficient volume computation, $\mathcal{R}_{c}$ is over-approximated by hyperrectangles. 
Denote $\mathcal{D} =  \left( e^{\bm{J}\Delta} \mathcal{R}^c(t-\Delta)  \right) \oplus  \int_0^{\Delta} e^{\bm{J}(\Delta-\tau)}\mathcal{W}_F \dd \tau \oplus \mathcal{R}_{el} $, which can be readily computed as the conventional reachset by \eqref{equ:rs:R point}.
Denote $\bm{E} = \int_0^\Delta e^{\bm{J}(\Delta-\tau)}\dd \tau$. 
Accordingly, the volume of the conformance reachset  is over-approximated as:

\begin{equation}
\begin{aligned}
    & \text{Vol}(\mathcal{R}_{c}(t)) = \text{Vol}( \mathcal{D} \oplus \bm{E}\mathcal{E}_m  ) \\
     \subseteq
    & \text{Vol}(\text{box}(\mathcal{D}) \oplus \bm{E}[\underline{\bm{e}}_m , \overline{\bm{e}}_m)]  )  \triangleq \text{Vol}(\overline{\mathcal{R}}_{c}(t))
\end{aligned}
\end{equation}

Consequently, the objective of  $\mathcal{E}_m$-opt, i.e., minimizing the conformance reachset volume, is simplified into minimizing the summation of edge lengths of the hyperrectangle $\overline{\mathcal{R}}_{c}(t)$. The new objective becomes:
\begin{equation} \label{equ:rsc:opt fnew}
    \min_{ \overline{\bm{e}}_m , \underline{\bm{e}}_m } \bm{1}^T (\sup{(D)}-\inf{(D)} + |\bm{E} | (\overline{\bm{e}}_m - \underline{\bm{e}}_m))
\end{equation}

On the other hand, the constraints of  $\mathcal{E}_m$-opt require that the NMs trajectories are enclosed by the conformance reachset $\mathcal{R}_c$, which are also handled by hyperrectangles:
\begin{equation} \label{equ:rsc:opt Cnew}
    \left\{
    \begin{aligned}
        & \hat{\bm{X}}(t) \leq \sup (\mathcal{D} \oplus \bm{E}\mathcal{E}_m ) =  \sup (\mathcal{D}) + \abs{\bm{E}} \overline{\bm{e}}_m   \\
        & \hat{\bm{X}}(t) \geq \inf (\mathcal{D} \oplus \bm{E}\mathcal{E}_m ) = \inf (\mathcal{D}) + \abs{\bm{E}} \underline{\bm{e}}_m
    \end{aligned}
    \right.
\end{equation}

Combining \eqref{equ:rsc:opt fnew} and \eqref{equ:rsc:opt Cnew} leads to a reformulated $\mathcal{E}_m$-opt model in the form of linear programming that can be tractably solved.

Further, with \textit{a set of time series} of NMs trajectories,  $\mathcal{E}_m$-opt results from each trajectory are joint to obtain the overall $\mathcal{E}_m$:
\begin{equation}
    \mathcal{E}_m =  \bigcup\nolimits_{j=1}^{m} \mathcal{E}_m^{(j)}
\end{equation}
where $\mathcal{E}_m^{(j)}$ denotes the optimization result from \eqref{equ:rsc:opt fnew} and \eqref{equ:rsc:opt Cnew} for the $j^{th}$ measurements.

\section{Case Studies}\label{sec:sim}
This section demonstrates the technical merit and efficacy of the \neuroFA method for NMs. 
The algorithm is implemented in MATLAB R2019b.

\subsection{Case Design}\label{sec:sim:case}

Case studies are conducted on the 4-microgrid NMs in Fig.~\ref{fig:main:NMs}, with the DER controller parameters modified from~\cite{pogaku2007modeling}. The DERs can be equipped with droop controllers or secondary controllers.
In the default scenario, the uncertainty of each DER is set as 20\%. 
Five cases are designed to verify the \neuroFA method:
\begin{itemize}[leftmargin=*]
\item[] \textbf{Case 1}: All the DERs are equipped with droop control;  microgrid 4  is supposed to be model-free and data-driven, while other microgrids are physics-based; 

\item[] \textbf{Case 2}: All the DERs, both in \insys and \exsys, are equipped with secondary control, and microgrid 4 is data-driven similar as Case 1;

\item[] \textbf{Case 3}: DERs are equipped with droop control, and microgrid 3 (which comprises 2 DERs) is data-driven;

\item[] \textbf{Case 4}: All settings are the same with Case 3, except that one DER in microgrid 3 is replaced by a synchronous generator (SG); 

\item[] \textbf{Case 5}: All settings are the same with Case 3, except that one DER in microgrid 3 is replaced by an energy storage unit (ESU); 
\end{itemize}

\subsection{ Validity of \ODENet-based NMs Model Discovery }
This subsection demonstrates the performance of \ODENet in learning the state-space model of microgrids. The trapezoidal rule is employed for ODE integration. The Adaptive Moment Estimation (Adam)~\cite{kingma2014adam} algorithm is applied to enable an adaptive learning rate during the \ODENet training.

\begin{figure}[!ht]
    \centering
    \begin{subfigure}{0.95\columnwidth}
        \begin{tikzpicture}
        \node[anchor=south west,inner sep=0] at (0,0)
        {\includegraphics[width=\columnwidth]{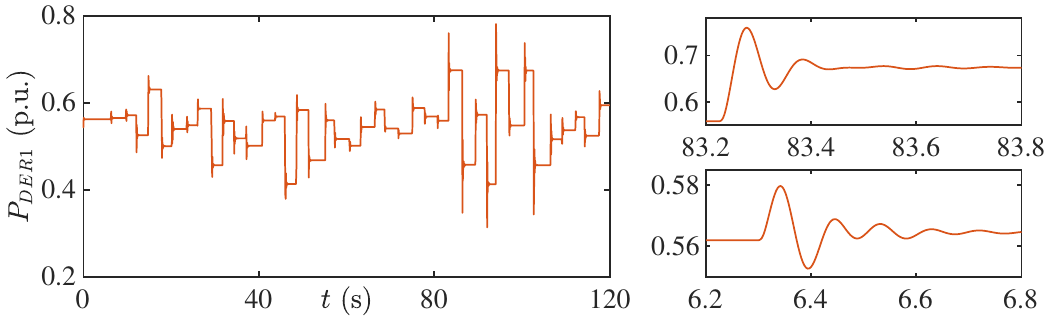}};
        \draw [line width=0.5] (0.9,1.7) ellipse (0.15 and 0.3);
        \draw[->, line width=0.5](1.0,1.4) -- (6.0,0.7);
        \node at (7.1,1) {\scriptsize \textcolor{black}{a small perturbation}};
        \draw [line width=0.5] (4.2,2.0) ellipse (0.15 and 0.5);
        \draw[->, line width=0.5](4.4,2.25) -- (5.8,2.25);
        \node at (7.1,2.25) {\scriptsize \textcolor{black}{a large perturbation}};
        \end{tikzpicture}
        \caption{Time-series measurements of NMs dynamics}
    \end{subfigure} \\
    \begin{subfigure}{0.325\columnwidth}
        \begin{tikzpicture}
        \node[anchor=south west,inner sep=0] at (0,0)
        {\includegraphics[width=\columnwidth]{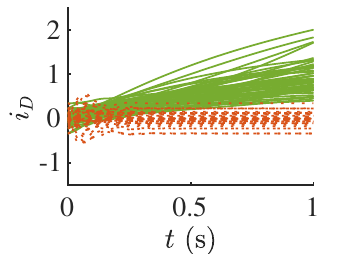}};
        \draw[->, line width=0.5](1.8,1.4) -- (1.6,1.8);
        \node at (1.6,2.0) {\scriptsize \textcolor{black}{Learned dynamics}};
        \draw[->, line width=0.5](1.8,1.1) -- (1.3,0.9);
        \node at (1.6,0.75) {\scriptsize \textcolor{black}{Real dynamics}};
        \end{tikzpicture}
        \caption{\ODENet performance at the starting stage}
    \end{subfigure}
    \begin{subfigure}{0.325\columnwidth}
        \includegraphics[width=\columnwidth]{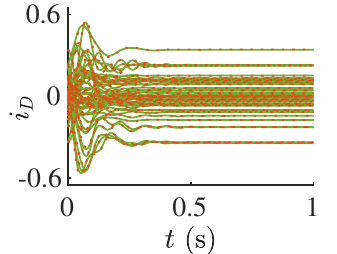}
        \caption{\ODENet performance at the final stage}
    \end{subfigure}
    \begin{subfigure}{0.325\columnwidth}
        \includegraphics[width=\columnwidth]{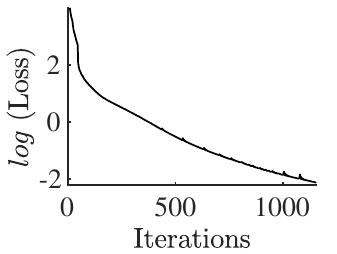}
        \caption{Loss function evolution process}
    \end{subfigure}
    \caption{Training process of \ODENet for NMs dynamic model discovery}
    \label{fig:sim:ODENet training}
    \vspace{-5pt}
\end{figure}
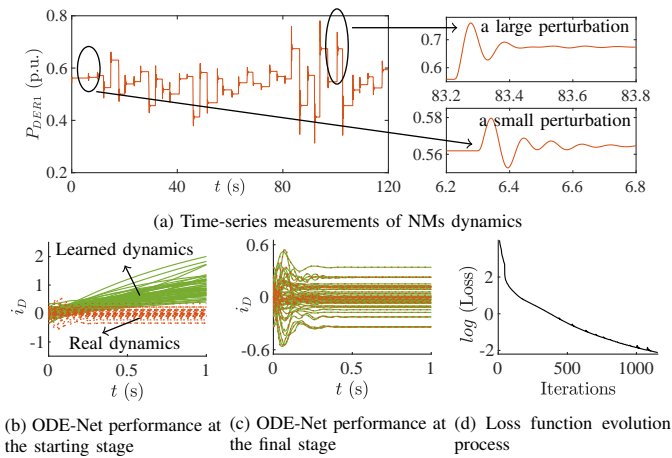

Fig.~\ref{fig:sim:ODENet training} presents the \ODENet training process  for Case 1.
As shown in Fig.~\ref{fig:sim:ODENet training}(a), the NMs undergo frequent transients due to the fluctuating of the DERs, which construct the training set for \ODENet learning.
At the starting stage, \ODENet is randomly initialized and largely deviates from the real NMs trajectories, as illustrated in Fig.~\ref{fig:sim:ODENet training}(b).
Then, after the neural network training via continuous backpropagation,  the \ODENet converges to a perfect match of the NMs trajectories on the training set, as illustrated in Fig.~\ref{fig:sim:ODENet training}(c). 
Additionally, Fig.~\ref{fig:sim:ODENet training}(d) presents the evolution of loss function at the logarithmic scale. As a rule of thumb, 1500 iterations lead to the convergence of the \ODENet.

\begin{figure}[!ht]
    \includegraphics[width=\columnwidth]{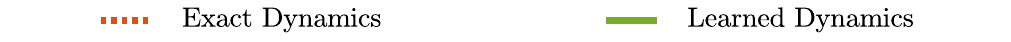}
    \centering
    \begin{subfigure}{0.325\columnwidth}
        \includegraphics[width=\columnwidth]{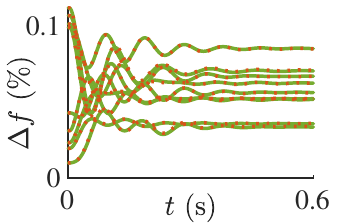}
        \caption{}
    \end{subfigure} 
    \begin{subfigure}{0.325\columnwidth}
        \includegraphics[width=\columnwidth]{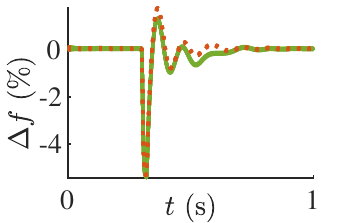}
        \caption{}
    \end{subfigure}
    \begin{subfigure}{0.325\columnwidth}
        \includegraphics[width=\columnwidth]{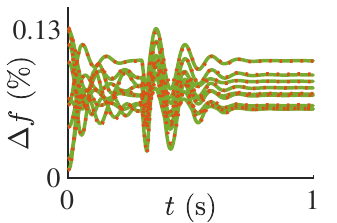}
        \caption{}
    \end{subfigure}\\
    \begin{subfigure}{0.325\columnwidth}
        \includegraphics[width=\columnwidth]{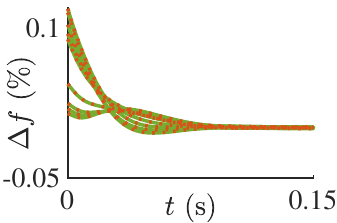}
        \caption{}
    \end{subfigure} 
    \begin{subfigure}{0.325\columnwidth}
        \includegraphics[width=\columnwidth]{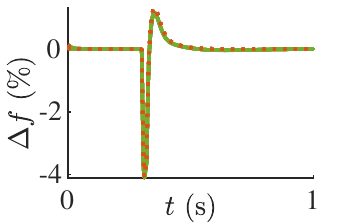}
        \caption{}
    \end{subfigure}
    \begin{subfigure}{0.325\columnwidth}
        \includegraphics[width=\columnwidth]{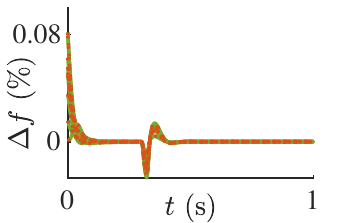}
        \caption{}
    \end{subfigure}\\
    \caption{\ODENet performance on the test set of NMs dynamics ($\Delta f$: frequency deviation). (a)-(c), droop control under no fault,  short-circuit fault and open-circuit fault respectively.  (d)-(f), secondary control under no fault,  short-circuit fault and open-circuit fault respectively.}
    \label{fig:sim:ODENet testing}
    \vspace{-5pt}
\end{figure}

Further, Fig.~\ref{fig:sim:ODENet testing} illustrates the \ODENet performance on the test set, which verifies its ability to generalize beyond the training set. 
Three types of scenarios are studied, i.e.,  no-fault, short-circuit fault, and open-circuit fault. 
A surprise finding is that the ODE-Net-enabled NMs formulation accurately captures the uncertain NMs transients not only under the frequently fluctuating DER uncertainties, but also under large disturbances, although the latter scenarios never appear in the training set. This shows the robustness of the ODE-net-enabled NMs formulation.

\vspace{-10pt}
\subsection{NMs Dynamic Verification Via \neuroFA}

This subsection studies the NMs dynamics with both uncertain perturbations and  fault disturbances via the \neuroFA analysis.

\begin{figure}[!ht]
    \centering
    \includegraphics[width= \columnwidth]{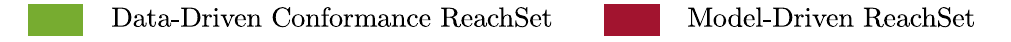}
    \vspace{5pt}
    \begin{subfigure}{\columnwidth}
        \includegraphics[width=0.48\columnwidth]{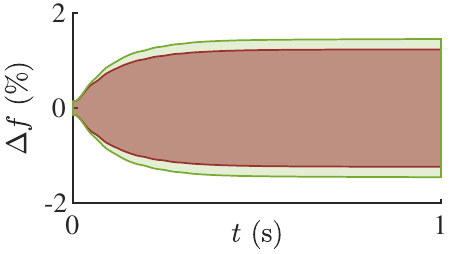}
        \includegraphics[width=0.48\columnwidth]{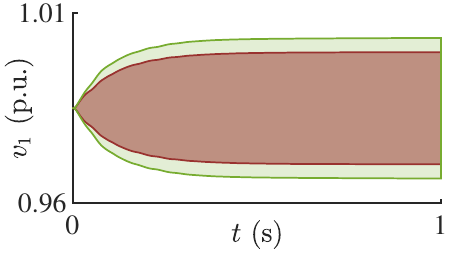}
        \caption{Case 1: DERs with droop control}
    \end{subfigure}
\\
    \begin{subfigure}{\columnwidth}
        \includegraphics[width=0.48\columnwidth]{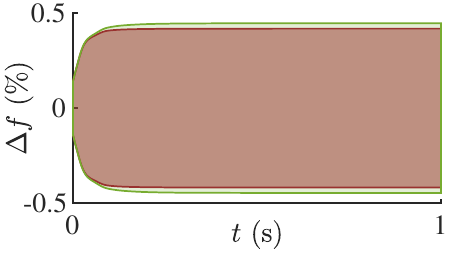}
        \includegraphics[width=0.48\columnwidth]{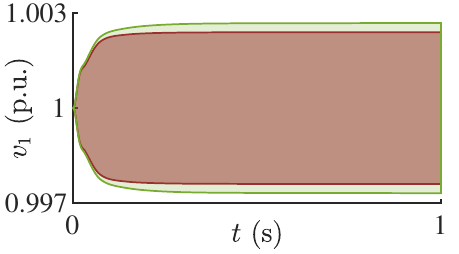}
        \caption{Case 2: DERs with secondary control}
    \end{subfigure}
   \caption{\neuroRScapital and its comparison with the model-driven reachsets ($\Delta f$: frequency deviation; $v_1$: voltage amplitude at DER1)}
   \label{fig:sim:case12 fault0}
   \vspace{-5pt}
\end{figure}

Fig.~\ref{fig:sim:case12 fault0} studies the \neuroRS under the quasi-static scenario, where the NMs is only perturbed by the DER uncertainties.
The simulation shows that the \neuroRS tightly encloses the model-driven reachsets, which verifies both the accuracy and conservativeness of the method. 
In particular, in Case 2 where the DERs are equipped with secondary control, the \neuroFA is supposed to learn not only the interactive currents between \exsys and \insys, but also the control signals from \exsys. Fig.~\ref{fig:sim:case12 fault0}(b) still shows a tight and perfect over-approximation of the real reachsets, which exhibits the potential of the \neuroFA method in learning the NMs dynamic model with complicated hierarchy control.
\neuroFA shows the superiority of secondary control from two aspects:
i) reachsets in Case 2 are narrower than the those in  Case 1, which reflects strengthened robustness of the NMs against uncertainties;
ii) reachsets in Case 2 get stable faster than those in  Case 1, which reflects a speedy power sharing between the DERs during the NMs dynamics induced by uncertain perturbations.

\begin{figure}[!ht]
    \centering
    \includegraphics[width= \columnwidth]{fig/legend_DDvsMD.pdf}
    \begin{subfigure}{\columnwidth}
        \begin{tikzpicture}
            \node[anchor=south west,inner sep=0] at (0,0) {\includegraphics[width=0.48\columnwidth]{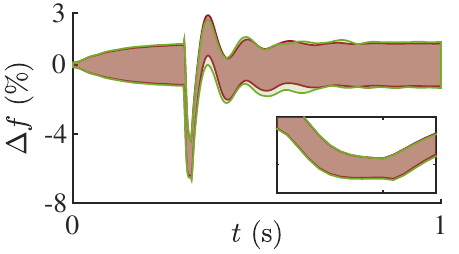}};
            \draw [line width=1 , black] (1.8,1.2) ellipse (0.3 and 0.6);
            \draw[->, line width=1 , black](2.1,0.9) -- (2.5,0.9);
        \end{tikzpicture}
        \begin{tikzpicture}
            \node[anchor=south west,inner sep=0] at (0,0) {\includegraphics[width=0.48\columnwidth]{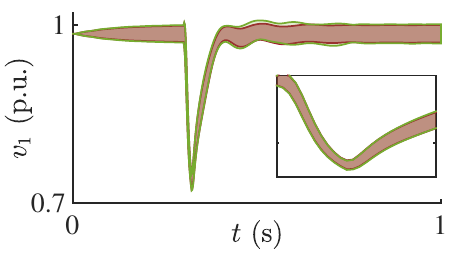}};
            \draw [line width=1 , black] (1.8,1.2) ellipse (0.3 and 0.6);
            \draw[->, line width=1 , black](2.1,0.9) -- (2.5,0.9);
        \end{tikzpicture}
        \caption{Case 1: DERs with droop control}
    \end{subfigure}
\\
     \begin{subfigure}{\columnwidth}
        \begin{tikzpicture}
            \node[anchor=south west,inner sep=0] at (0,0) {\includegraphics[width=0.48\columnwidth]{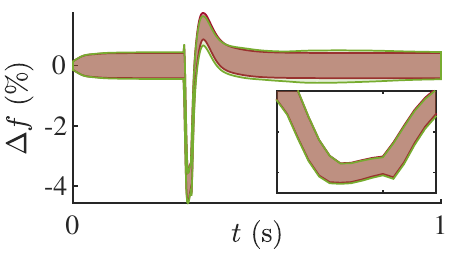}};
            \draw [line width=1 , black] (1.8,1.2) ellipse (0.3 and 0.6);
            \draw[->, line width=1 , black](2.1,0.9) -- (2.5,0.9);
        \end{tikzpicture}
        \begin{tikzpicture}
            \node[anchor=south west,inner sep=0] at (0,0) {\includegraphics[width=0.48\columnwidth]{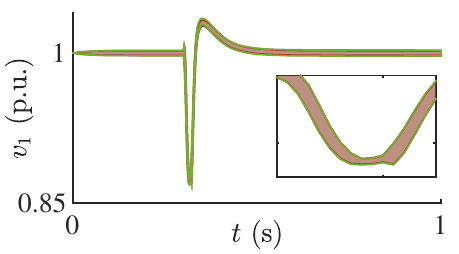}};
            \draw [line width=1 , black] (1.8,1.2) ellipse (0.3 and 0.6);
            \draw[->, line width=1 , black](2.1,0.9) -- (2.5,0.9);
        \end{tikzpicture}
        \caption{Case 2: DERs with secondary control}
    \end{subfigure}
   \caption{\neuroRScapital under a short-circuit fault}
    \label{fig:sim:case12 fault1}
    \vspace{-5pt}
\end{figure}

Further, Fig.~\ref{fig:sim:case12 fault1} studies the \neuroRS under a momentary  short-circuit fault occurred at 0.3s and cleared at 0.32s. 
Even though the \ODENet-based NMs formulation is learned from the dynamics of small disturbances induced by the DERs' uncertainties, simulation results show that the \neuroFA method accurately and tightly captures the fast NMs dynamics under heterogeneous uncertainties during disturbances. 
Comparing the \neuroRS in Case 1 and Case 2, it is obvious that the secondary control exhibits restrained frequency/voltage dips and enhanced damping on the oscillations of the NMs states.
Specifically, the reachable sets  during the periods from the fault occurrence through the fault clearance are magnified by the subplots. \neuroFA thus effectively mimics the NMs transients initiated by large disturbances.

\vspace{-10pt}
\subsection{Efficacy and Versatility of \neuroFA}
This subsection studies the impact factors of the NMs reachsets via the \neuroFA method.

\subsubsection{\neuroRSCAPital Under Different Uncertainty Levels}

\begin{figure}[!ht]
\vspace{-5pt}
    \centering
    \includegraphics[width=\columnwidth]{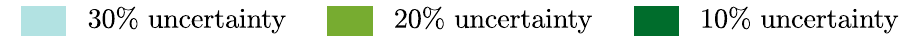} \\
    \vspace{5pt}
    \begin{subfigure}{0.48\columnwidth}
        \includegraphics[width=\columnwidth]{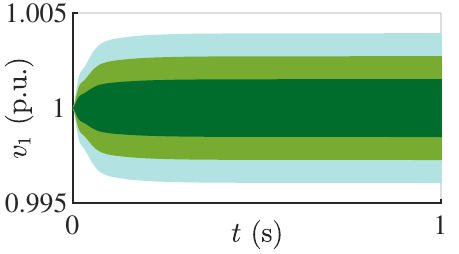}
        \caption{No fault}
    \end{subfigure}
    \begin{subfigure}{0.48\columnwidth}
        \includegraphics[width=\columnwidth]{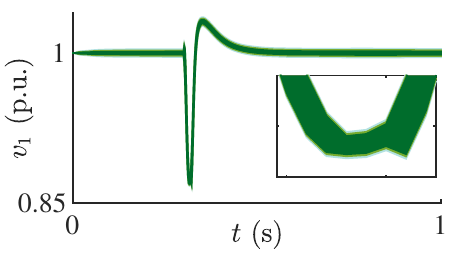}
        \caption{Short-circuit fault}
    \end{subfigure}
    \caption{\neuroRScapital under different uncertainty levels of Case 2}
    \label{fig:sim:uncertainty}
    \vspace{-5pt}
\end{figure}

First, the impact of uncertainties on NMs dynamics is investigated.
Taking DER1's output-voltage in Case 2 as an example, Fig.~\ref{fig:sim:uncertainty} shows that the reachsets expand with the increasing uncertainties under both no-fault and faulted scenarios. 
The propagation of uncertainties in the NMs dynamics is therefore distinctly demonstrated by the \neuroRS at different uncertainty levels.

\subsubsection{\neuroRSCAPital Under Different Control Strategies}
\begin{figure}[!ht]
\vspace{-5pt}
    \centering
    \begin{subfigure}{0.48\columnwidth}
        \includegraphics[width=\columnwidth]{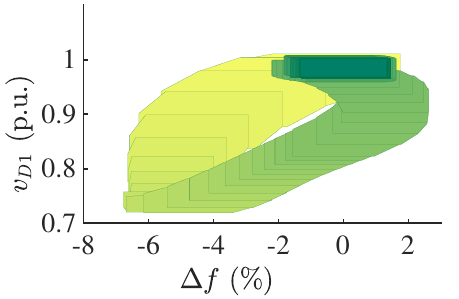}
        \caption{Case 1: DERs with droop control}
    \end{subfigure}
    \begin{subfigure}{0.48\columnwidth}
        \includegraphics[width=\columnwidth]{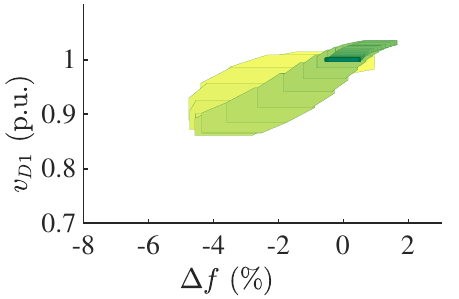}
        \caption{Case 2: DERs with secondary control}
    \end{subfigure}
    \caption{State space \neuroRS under different control strategies}
    \label{fig:sim:control}
    \vspace{-5pt}
\end{figure}

Second, the impact of the DERs' control strategies on NMs reachsets is investigated. Fig.~\ref{fig:sim:case12 fault0} and Fig.~\ref{fig:sim:case12 fault1} exhibit the efficacy of the secondary control to stabilize the system under uncertainties from the time-domain perspective. Further, Fig.~\ref{fig:sim:control} presents the \neuroRS from a state-space observation. The efficacy of the secondary control for restraining the uncertainty impacts,  damping the frequency/voltage overshoot and recovering the NMs states  after faults are distinctly verified via the \neuroRS.
Hence, \neuroFA is promisingly helpful for verifying the controller's performance under the infinite many uncertain scenarios in the absence of the microgrid model.

\subsubsection{\neuroRSCAPital Under Different Power Source Mixes}
Finally, the impact of power source mixes is investigated via \neuroRS. Different power sources influence the NMs transients by their diverse dynamic features.

\begin{figure}[!ht]
\vspace{-5pt}
    \centering
    \includegraphics[width= \columnwidth]{fig/legend_DDvsMD.pdf}
    \begin{subfigure}{\columnwidth}
        \begin{tikzpicture}
        \node[anchor=south west,inner sep=0] at (0,0)
        {\includegraphics[width=0.32\columnwidth]{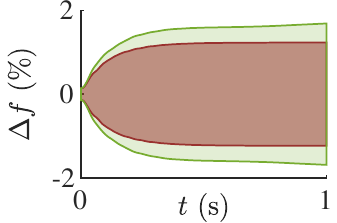}
        \includegraphics[width=0.32\columnwidth]{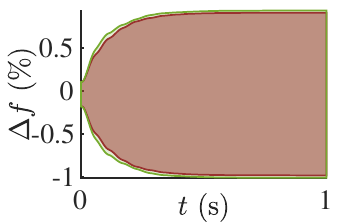}
        \includegraphics[width=0.32\columnwidth]{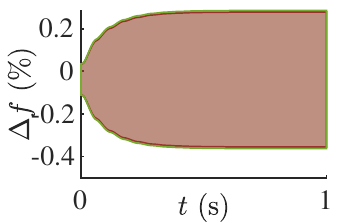}};
        \node at (1.6,2.1) {\scriptsize \textcolor{black}{Case 3: 2 DERs}};
        \node at (4.7,2.1) {\scriptsize \textcolor{black}{Case 4: 1DER + 1SG}};
        \node at (7.6,2.1) {\scriptsize \textcolor{black}{Case 5: 1DER + 1ESU}};
        \end{tikzpicture}
        \caption{Frequency reachset}
    \end{subfigure} 
    \vspace{5pt}
    \begin{subfigure}{\columnwidth}
        \begin{tikzpicture}
        \node[anchor=south west,inner sep=0] at (0,0)
        {\includegraphics[width=0.32\columnwidth]{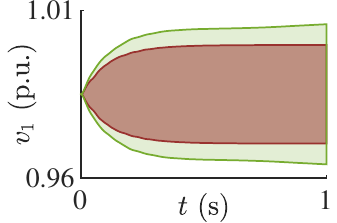}
        \includegraphics[width=0.32\columnwidth]{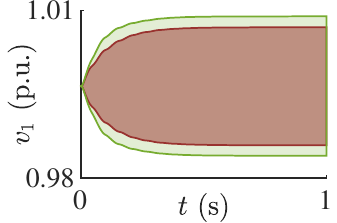}
        \includegraphics[width=0.32\columnwidth]{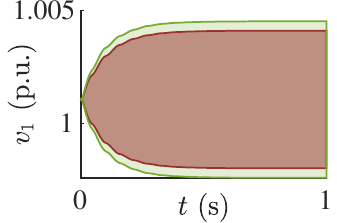}};
        \node at (1.6,2.1) {\scriptsize \textcolor{black}{Case 3: 2 DERs}};
        \node at (4.7,2.1) {\scriptsize \textcolor{black}{Case 4: 1DER + 1SG}};
        \node at (7.6,2.1) {\scriptsize \textcolor{black}{Case 5: 1DER + 1ESU}};
        \end{tikzpicture}
        \caption{Voltage reachset}
    \end{subfigure} 
    \caption{\neuroRScapital under a short-circuit fault under different mixes of power sources in NMs}
    \label{fig:sim:source1}
    \vspace{-5pt}
\end{figure}

\begin{figure}[!ht]
\vspace{-5pt}
    \centering
    \begin{subfigure}{0.48\columnwidth}
        \includegraphics[width=\columnwidth]{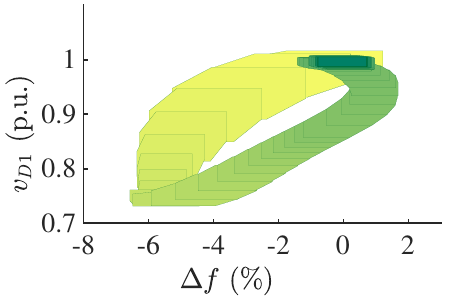}
        \caption{Case 4}
    \end{subfigure}
    \begin{subfigure}{0.48\columnwidth}
        \includegraphics[width=\columnwidth]{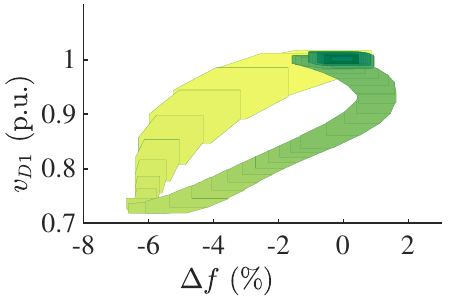}
        \caption{Case 5}
    \end{subfigure}
    \caption{\neuroRScapital under different mixes of power sources in NMs}
    \label{fig:sim:source2}
    \vspace{-5pt}
\end{figure}

Fig.~\ref{fig:sim:source1} studies the quasi-static reachsets (i.e., with no fault but only DER uncertainties perturbing the NMs) for Case 3, Case 4 and Case 5. The \neuroRS tightly encloses the model-driven reachsets in all cases, which again verifies the correctness of the \neuroFA method. 
An interesting finding is that the \neuroRS of Case 3 in Fig.~\ref{fig:sim:source1} are nearly identical to those of Case 1 in Fig.~\ref{fig:sim:case12 fault0}. This is  because case 1 and Case 3 describe the identical NMs only with different microgrids being data-driven. 
Results of Case 4, compared with those of Case 3, show that the \neuroRS shrinks when the NMs are equipped with a SG, benefiting from the inertia and regulating ability of SG as well as its full dispachability compared with DERs.  
The \neuroRS is further improved when the NMs are equipped with an ESU as presented by the \neuroRS of Case 5, indicating enhanced robustness against the uncertainties with ESU.
Meanwhile, it is noteworthy that both the SG and ESU provide voltage support to boost the NMs voltage, as illustrated in Fig.~\ref{fig:sim:source1}(b). 
Additionally, Fig.~\ref{fig:sim:source2} investigates the \neuroRS of NMs during a short-circuit fault for Case 4 (i.e., equipped with a SG) and Case 5 (i.e., equipped with an ESU). The \neuroFA method, as a data-driven approach, successfully captures the transient characteristics of the controllable power sources in restraining the uncertainties and damping the frequency/voltage dip/rise during the large disturbances compared with Fig.~\ref{fig:sim:control}(a).

\vspace{3pt}

\section{Conclusions} \label{sec:conclusion}
This paper devises a \neuroFA method, a data-driven approach  to verify the uncertainty-disturbed, fast-changing and strongly-nonlinear dynamics of the NMs with unidentified subsystems. 
The \ODENet-enabled dynamic model discovery, reachability analysis, and conformance theory conjointly enable a flexible and accurate model discovery of real-world microgrids, as well as provide reliable reachsets for verifying the NMs dynamics under heterogeneous uncertainties.  
Case studies of a typical NMs demonstrate the efficacy and robustness of the devised method. 
In the future, the \neuroFA method will be enhanced for the dynamic verification of fully model-free NMs. 

\vspace{3pt}

\bibliographystyle{ieeetr}

\end{document}